\documentclass[aps,showpacs,prl,superscriptaddress]{revtex4}
\usepackage[utf8]{inputenc}
\usepackage{graphicx}
\usepackage{amssymb}
\usepackage{epstopdf}
\usepackage{color, amsmath, bm}
\usepackage{ulem}
\usepackage[switch, modulo]{lineno} 
\usepackage[hidelinks]{hyperref}
\newcommand{\be}{\begin{equation}}
\newcommand{\ee}{  \end{equation}}
\newcommand{\ba}{\begin{eqnarray}}
\newcommand{\ea}{  \end{eqnarray}}
\newcommand{\ket}[1]{\left|#1\right>}
\newcommand{\bra}[1]{\left< #1 \right|}
\newcommand{\braket}[2]{\left< #1| #2 \right>}

\begin{document}

\title{Towards detecting traces of non-contact quantum friction in the corrections of the accumulated geometric phase}
\author{M. Bel\'en Far\'\i as}
\affiliation{Departamento de F\'\i sica {\it Juan Jos\'e
Giambiagi}, FCEyN UBA and IFIBA CONICET-UBA, Facultad de Ciencias Exactas y Naturales,
Ciudad Universitaria, Pabell\' on I, 1428 Buenos Aires, Argentina.}
\affiliation{University of Luxembourg, Physics and Materials Science Research Unit, Avenue de la Fraïncerie 162a, L-1511, Luxembourg, Luxembourg}
\author{Fernando C. Lombardo$^*$}
\affiliation{Departamento de F\'\i sica {\it Juan Jos\'e
Giambiagi}, FCEyN UBA and IFIBA CONICET-UBA, Facultad de Ciencias Exactas y Naturales,
Ciudad Universitaria, Pabell\' on I, 1428 Buenos Aires, Argentina.}
\author{Alejandro Soba}
\affiliation{ Centro  At\'omico  Constituyentes,  Comisi\'on  Nacional  de  Energ\'\i a  At\'omica,
Avenida  General  Paz  1499,  San  Mart\'\i n,  Argentina}
\author{Paula I. Villar}
\affiliation{Departamento de F\'\i sica {\it Juan Jos\'e
Giambiagi}, FCEyN UBA and IFIBA CONICET-UBA, Facultad de Ciencias Exactas y Naturales,
Ciudad Universitaria, Pabell\' on I, 1428 Buenos Aires, Argentina.}
\author{Ricardo S. Decca}
\affiliation{Department of Physics, Indiana University-Purdue University Indianapolis, Indianapolis, Indiana 46202, USA. $^*$email: lombardo@df.uba.ar}
\date{\today}                                           

\begin{abstract}
\noindent 
The geometric phase can be used as a fruitful venue of investigation to infer features of the quantum systems. Its application can reach new theoretical frontiers and imply innovative and challenging experimental proposals. Herein,  we take advantage of the geometric phase to sense the corrections induced while a neutral particle travels at constant velocity in front of an imperfect sheet in quantum vacuum. As it is already known, two bodies in relative motion at constant velocity experience a quantum contactless dissipative force, known as quantum friction. This force has eluded experimental detection so far due to its small magnitude and short range. However, we give details of an innovative experiment designed to  track traces of the quantum friction by measuring the velocity dependence of  corrections to the geometric phase. We notice that the environmentally induced corrections can be decomposed in different contributions: corrections induced by the presence of the dielectric sheet and the motion of the particle in quantum vacuum.  As the geometric phase accumulates over time, its correction  becomes relevant at a relative short timescale, while the system still preserves purity. The experimentally viable scheme presented would be the first one  in tracking traces of quantum friction through the study of decoherence effects on a NV center in diamond.
\end{abstract}

\maketitle
\noindent \textbf{INTRODUCTION} \\

\begin{figure*}
\centering
\includegraphics[width=\textwidth]{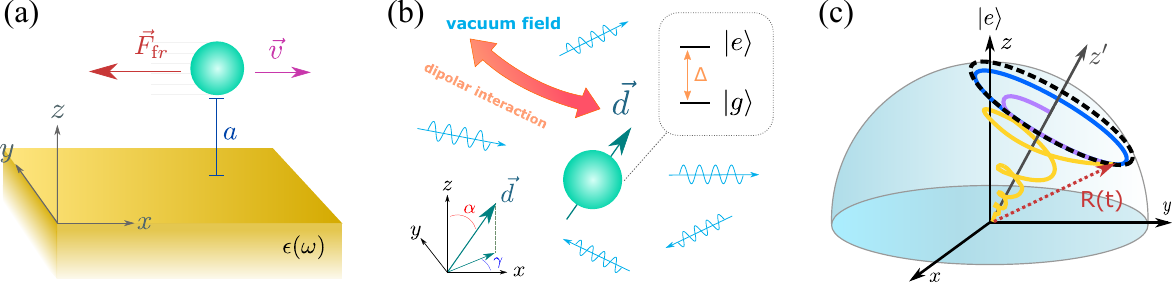}
\caption{We study the GP acquired by a two-level system coupled to the quantum vacuum in front of a dielectric sheet. (a)  The center of mass of the particle moves with constant velocity ${\bf v} $ at a fixed distance $a$ of the dielectric sheet. (b) The particle behaves as a two-level system coupled via a dipolar interaction to the vacuum field, which is \textit{dressed} by the presence of the dielectric plate. (c) Scheme of the evolution of a two-level system in the Bloch sphere for an unitary evolution (black dashed line) and environmentally induced trajectories. The unitary (closed) geometric phase is $\phi_c=\pi(1-\cos\theta_0)$, while in the presence of an environment the area enclosed is modified, yielding under suitable conditions $\phi_g=\phi_c +\delta\phi$, where we define $\delta \phi$ as the correction to the unitary geometric phase. In all cases, the presence of the environment induces a loss of purity which is represented in the fact that the trajectory leaves the surface of the sphere or semi-sphere.}
\label{Fig1}
 \end{figure*}

\noindent One of the most exciting features of quantum field theory is based on the 
nontrivial structure of the vacuum state and the observable macroscopic effects associated to quantum vacuum fluctuations.
The most renowned example is the attractive Casimir force between two neutral bodies \cite{Casimir,Milonni,Bordag1,Bordag2,Milton1,Milton2,Reynaud,Lamoreaux}. There is another fascinating effect when a mirror moves through space at relativistic speeds: 
some photons become separated from their partners and the mirror begins to produce light. This phenomenon is known as dynamical Casimir effect (DCE) \cite{DCE1,DCE2}. 
Likewise, it has been recently shown that two bodies moving relative to each other at constant velocity experience a dissipative force that opposes the motion due to the exchange of Doppler shifted virtual photons, known as quantum friction (QF)  \cite{Pendry97,debate1,debate2,debate3,debate4,vp2007}. So far, the static Casimir force has been measured \cite{Mohideen1,Mohideen2,Mohideen3,Mohideen4,Mohideen5,Mohideen6}. However, the challenge of making a mirror or neutral particle to move at an almost relativistic speed in a medium has prevented a direct experimental observation of such dynamical effects. As a result, the only dynamic-observable macroscopic effect that has allowed experimental observation of DCE was based on  electromagnetic analogs of a moving mirror using a tunable reflecting element in a superconducting device \cite{Wilson,Lahteenmaki}. Similarly, since  the predicted QF is very small in magnitude and short ranged, its experimental detection has become an absolute challenge so far \cite{Gotsmann}.  Lately, there have been a variety configurations \cite{spheres} and theoretical efforts devoted into finding favorable conditions for experimental measurements of QF \cite{Farias1,Farias2,Farias3,intra,Fosco1,Fosco2,Buhmann}. In Refs. \cite{Volokitin,vpPRB}, authors have investigated the van der Waals friction between graphene and an amorphous SiO$_2$ substrate. They found that due to this friction the electric current is saturated at a high electric field. The saturation current depends weakly on the temperature, which they attributed to the quantum friction between the graphene carriers and the substrate optical phonons. They calculated the 
frictional drag between two graphene sheets caused by van der Waals friction, and proved that this drag can induce a voltage high enough to be measured experimentally 
by state-of-art non-contact force microscopy. This work paved the way for a possible mechanical detection of the Casimir friction. \\

\noindent In a different approach, in this article we propose to track traces of QF through the measurement of the geometric phase (GP) accumulated by an atom moving at constant velocity in front a dielectric material in vacuum. The GP shift is manifested as a relative phase between components of a superposition of atomic states. In \cite{EPL} some of us have proposed a very simplistic model with this idea, where a harmonic oscillator (the ``atom") is coupled to an scalar field (as the ``vacuum field") and the later is 
coupled to a mirror composed by a set of harmonic oscillators without dissipation and with an unique natural frequency.  Therein the coupling considered was purely dephasing  just for the sake of simplicity, far away of any real experimental consideration. We have even evaluated an unique decoherence factor (neglecting other noise sources) in two spacial dimensions, under some strong approximations.
 
\noindent The state of a point like discrete energy level quantum system (such as an atom) interacting with a quantum field acquires a GP that is independent of the state of the field \cite{Berry}. The phase depends only on the system's path in parameter space, particularly the flux of some gauge field enclosed by that path. For pure field states, the GP is said to encode information about the number of particles in the field \cite{Fuentes1}. In particular, for initial squeezed states, the phase also depends on the squeezing strength \cite{Fuentes2}. If the field is in a thermal state, the GP encodes information about its temperature, and so is used in a proposal to measure the Unruh effect at low accelerations \cite{Martin}.  
It has further been proposed as a high-precision thermometer by considering the atomic interference of two atoms interacting with a known hot source and an unknown temperature cold cavity \cite{Martin2}. As the existing bibliography reflects, GPs have become a fruitful avenue of investigation to infer features of the quantum system due to their topological properties and close connection with gauge theories of quantum fields.
Berry's work has since been applied to a diversity of phenomena, extending its definition to mixed states \cite{Sjoqvist}, non-adiabatic \cite{Moore} and non-unitary evolution \cite{Tong1,Tong2} in the context of open quantum systems, largely motivated by the need of an experimentally realistic definition. 
Every  two-level system freely  evolving acquires an unitary GP related to its Hilbert Space, i.e. $SU(2)$, known as $\phi_c=\pi(1-\cos\theta_0)$, being $\theta_0$ the polar angle in the Bloch sphere. With this fact in mind, our idea is to study the GP acquired by the quantum system evolving non-unitarily.  The GP obtained will undoubtedly be different to the unitary one since the evolution is now plagued by non-unitary effects such as decoherence and dissipation. It is generally said that the coupling of the quantum system corrects the unitary GP by noting that $\phi_g=\phi_c +\delta\phi $, being $\delta\phi$ proportional to the coupling of the system and the environment. Under suitable conditions, the corrections induced by the presence of the environment can be measured by means of an interferometric experiment (atomic  interference) 
\cite{Zeilinger,Leek,rotating1,rotating2}.  In this framework, we shall make a thorough study of the corrections  induced on the GP acquired by a neutral particle moving in front of an imperfect mirror.\\

\noindent \textbf{RESULTS}\\

\noindent \textbf{The Model}\\

\noindent We shall consider a neutral particle coupled to a vacuum
field, whose center of mass moves with a velocity $\mathbf v$ relative to the dielectric as schematically shown in Fig. \ref{Fig1}(a).  
The radiation emitted by the moving particle leads to a frictional stress acting on it, due to the
asymmetry of the reflection amplitude along the direction of motion \cite{Volokitin}. The particle is modeled as a two-level system and its velocity is assumed to be rendered constant by some external agent.
The total Hamiltonian for the system (quantum particle plus complex environment - composed by the vacuum quantum field and dielectric plate)
is written as
$H= \hbar/2 ~ \Delta ~\sigma_z + H_{\cal S E} + H_{\cal E}$,
where $\Delta$ is the energy level spacing of the two-level system 
and $H_{\cal E}$ is the Hamiltonian of the composite bath.
The interaction Hamiltonian $H_{\cal S E}$ is given in the dipolar approximation  $H_{\cal S E}= \hat{\mathbf d} \cdot \bm\nabla \Phi$, 
where  the effective electrostatic potential ($\Phi$) contains the information of the electric field ($\mathbf E = - \bm\nabla \Phi$) dressed by the 
dielectric surface.  The dipole operator $\hat{\mathbf{d}}$ acts over the internal states of the particle $\ket{g}$ and $\ket{e}$, and the non-vanishing matrix elements of each component are 
given as $d_i = \langle g\vert \hat{d}_i\vert e\rangle = \langle e\vert \hat{d}_i\vert g\rangle$ ($\sigma_x$ coupling). An schematic picture of the system is depicted in Fig. \ref{Fig1}(b), where the particle interacts with the dressed vacuum field. \\

\noindent The field operator $\hat{\Phi}$ can be expanded in a plane-wave basis of excitation (dressed photons) and evolves freely 
according to 
\begin{equation} \hat{\Phi}(\mathbf r, t) = \int d^2k \int_0^\infty d\omega \left(  \phi(\mathbf{k},\omega) \hat{a}_{\mathbf k,\omega} e^{i (\mathbf k \cdot \mathbf r - \omega t)} 
+ h.c.\right). \nonumber \end{equation}
The bosonic operators satisfy the commutation relation $[\hat{a}_{\mathbf{k}\omega},  \hat{a}^\dagger_{\mathbf{ k}'\omega'} ] = 
\delta(\mathbf{k} - \mathbf{k}') \delta(\omega - \omega')$  creating and destroying  ``photons"  in a wider meaning, since 
they are creation and destruction operators of composite states (field plus material). In particular, it has been shown that \cite{dalvitNM} 
\begin{equation}
\vert \phi(\mathbf{k},\omega)\vert^2 = \frac{\hbar}{2\pi^2} \frac{e^{-2kz}}{k} {\mbox{Im}}
\frac{\epsilon(\omega) - 1}{\epsilon(\omega) +1} \,,
\end{equation}
where $\epsilon(\omega)$ is the dielectric function of the dielectric material $\epsilon(\omega) = \omega_{\rm pl}^2/(\omega_0^2 - \omega^2 - i \omega\Gamma)$. In the dielectric model $\omega_0$, $\omega_{\rm pl}$ and $\Gamma$ parametrize the 
Drude-Lorentz form for $\epsilon(\omega)$. Here, $\omega_0^2 = \omega_S^2 - \omega_{\rm pl}^2/2$ 
with $\omega_S$ 
the surface-plasmon frequency, $\omega_{\rm pl}$ is the plasma frequency and $\omega_0=0$ represents the Drude model for 
metals. We consider only weak dissipation, i.e. $\Gamma/\omega_S \ll 1$; in metals, this ratio is typically $10^{-2} - 10^{-3}$ \cite{barton}. \\

\noindent \textbf{Non-unitary evolution of the system}\\

\noindent In order to get an insight into the dynamics of the quantum system at a later time $t$, we must obtain the master equation for the reduced density matrix of the quantum system derived by 
integrating out the degrees of freedom of the composite environment. 
We start by assuming an initially factorized state 
$\rho(0)= \rho_a(0) \otimes \rho_{\cal E }(0)$, with both sub-systems (particle and environment) initially in their respective 
ground states. In the  weak coupling limit, the master equation is given by the expression \cite{mastereq,mastereq2},
\ba
\hbar \dot\rho &=& -i \left[H_a, \rho\right] - D({\bf v},t) \left[\sigma_x,\left[\sigma_x,\rho\right]\right]  \label{master} \\
&&- f({\bf v},t) \left[\sigma_x,\left[\sigma_y,\rho\right]\right] 
+  i \zeta ({\bf v},t)\left[\sigma_x,\left\{\sigma_y,\rho\right\}\right] , \nonumber
\ea where the non-unitary effects are modeled by the velocity dependent diffusion coefficients $D({\bf v},t)$ and $f({\bf v},t)$, while 
dissipative effects are present in the corresponding $\zeta ({\bf v},t)$ \cite{refs1,refs2,refs3,pra1,pra2,pra3,pra4,pra5,pra6,prl}.
 We set the dimensionless quantity ${\bf u} = {\bf v}/(a \omega_{\rm pl})$, which will be called the velocity from here on. 
It is important to note that in absence of the bath,
the spin follows a cyclic evolution in the Bloch semi-sphere with period $\tau=2\pi/\Delta$, as can be seen 
in Fig.\ref{Fig1}(c) represented by the dashed-black line. \\

\noindent We have numerically calculated the coefficients in equations (\ref{master}) by using the expressions obtained in equation (\ref{noisedisspkernels}) (defined in section Materials and Methods). These 
coefficients are oscillating functions of time, strongly dependent on the parameters of the model. We shall show 
the case of metallic ($\omega_0 = 0$) mirrors in order to solve the master equation (\ref{master}). We prepare an initial superposition state as $|\Psi_0\rangle = \cos(\theta_0) |g\rangle+  \sin(\theta_0)  |e\rangle$ and solve the master equation to find the state of the system at a later time,  $|\Psi (t) \rangle = \cos(\theta(t)) |g\rangle +  \sin(\theta(t))  |e\rangle$. For a ground-state atom the force is a genuine friction force, i.e. a force antiparallel to its velocity. It is proportional to the atomic linewidth and hence very small. In contrast, excited-state atoms can be either decelerated or accelerated depending on the relative magnitude of their transition frequency with respect to the characteristic frequency of the substrate material. For metals, the force is always decelerating, i.e. a frictional effect \cite{buhmannscheel}. \\

\begin{figure*}[t]
	\centering
	\includegraphics[width=0.97\textwidth]{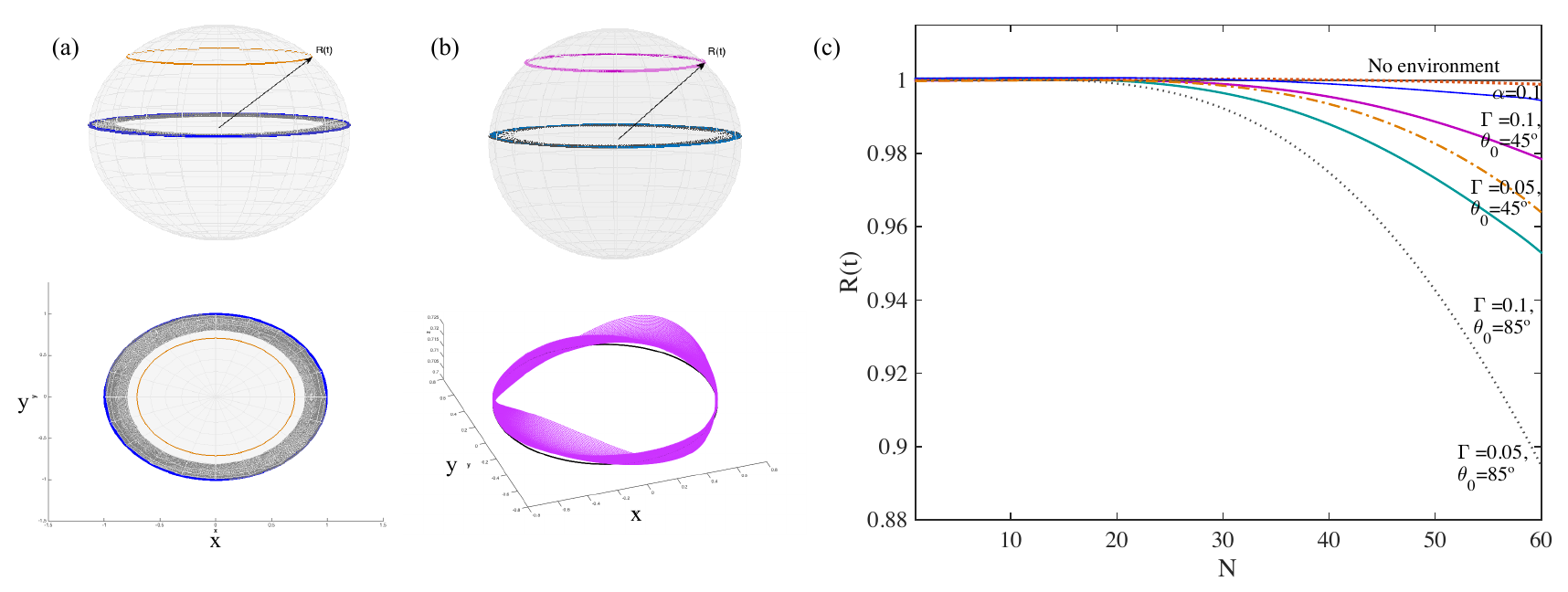}
	\caption{Non-unitary behaviour of the two-level system. In (a) and (b) we present different trajectories of the state vector in the Bloch sphere from a side and upper point of view. Upper spheres show trajectories for different initial superposition states ($\theta_0=45^{\circ}$ and $\theta_0=85^{\circ}$), while the polar and side view tries to get an insight into the trajectory itself. In (a), the lower polar view we can see the loss of purity for an environment induced evolution at $\theta_0=85^\circ$, as the state leaves the initial surface of the sphere.  The black curve represents $\Gamma/\omega_{\rm pl}=0.05$ and $\alpha=\pi/2$  while the dark blue curve is for $\Gamma/\omega_{\rm pl}=0.1$ and $\alpha=0.1$. The isolated evolution is represented by an orange line at $\theta_0=45^\circ$ for reference. In (b), we  show an environment-induced evolution at  $\theta_0=45^\circ$ ($\Gamma/\omega_{\rm pl}=0.1$ and $\alpha=\pi/2$) with a pink line, while for  $\theta_0=85^\circ$, we show the differences in varying $\Gamma$: i.e dark green curve ($\Gamma/\omega_{\rm pl}=0.1$ and $\alpha=\pi/2$) and black curve ($\Gamma/\omega_{\rm pl}=0.05$ and $\alpha=\pi/2$). Lower side view is an insight into the evolution of the pink line (compared to the isolated trajectory). (c) Loss of purity as a function of time (cycles) for the different trajectories shown. Other parameters used are: $\mathrm u = v_x/(a\omega_{\rm pl}) = 0.007$, $\Delta/\omega_{\rm pl}= 0.9$, $\omega_0/\omega_{\rm pl}=0$, and $\gamma = 0$.}
	\label{Fig2}
\end{figure*}

\noindent
In Fig. \ref{Fig2}(a) we present the trajectory of the radius of the state vector $R=\sqrt{x(t)^2+y(t)^2+z(t)^2}$ on the Bloch sphere when the system evolves
coupled to different values for the parameters of the environment. The trajectory of the state vector remains on the surface of the sphere for many cycles although the dynamics is very different to that of a free two-level system (a perfect black circle), as shown in the lower  part of Fig. \ref{Fig2}(b). The variation of the radius and its sinusoidal movement are induced by the environment and evince a non-unitary evolution.
The advantage of this system is that it remains ``robust" under the presence of decoherence and dissipation even after several cycles. The measurement of this ``robustness" is that the loss of purity of the state vector  is very small  for approximately 30 cycles.
The dependence of this magnitude upon time (measured in cycles) depends on the parameters of the model as shown in Fig. \ref{Fig2}(c). As it can be seen there, the state is more affected for smaller values of $\Gamma$ and bigger initial angles $\theta_0$, as long as it has a dipole orientation parallel to the dielectric sheet.
The purity of the state remains close to unity (isolated case) when the dipole orientation is almost perpendicular to the dielectric sheet.
These considerations are made based on a fixed velocity.
The solution renders a good scenario for measurements of the GP, since it has been argued that the observation of GPs should be done in times long enough to obey the adiabatic 
approximation but short enough to prevent decoherence from deleting all phase information. This means that while there are dissipative and diffusive effects that induce a correction to the
unitary GP,
the system maintains its purity for several cycles, which allows the GP to be observed. It is important to note that if the noise effects induced on the system
are of considerable magnitude, the coherence terms of the quantum system are rapidly destroyed and the GP literally disappears \cite{pra1}.\\

\noindent \textbf{Traces of Quantum Friction} \\

\noindent If the two-level system was to evolve freely, it would acquire a global phase at a temporal step $\tau$. The presence of an environment modifies this scenario. As we have seen above, the dynamics of the system differs from the isolated evolution due to noise effects. The notion of GP in the context of quantum open systems has been obtained in \cite{Tong1} by the use of a kinematical approach and the reduced density matrix. 
This definition satisfies the condition of being gauge independent and reproduces the already known result of GP in the isolated case, i.e. when the system is not coupled to an environment and the evolution is unitary. It has been extensively used to measure the corrections of the GP in a non-unitary evolution \cite{pra1,prl} and explain the noise effects in the observation of the GP in a superconducting qubit \cite{Leek,nosqubit}. 
In order to compute the GP we assume an initial state of the two-level system $|\Psi_0\rangle$ and solve the master equation to find the state at a later system $|\Psi (t) \rangle$. Once the state is known, we can build the reduced density matrix $\rho(t)=|\Psi(t)\rangle\langle \Psi(t)|$.
By computing the eigenvalues and eigenvectors of $\rho(t)$ we can obtain the GP as a function of time.
In order to compare with the unitary GP, we shall compute the open GP in cycles defined as $\rm N= t/ \tau$. We study the GP for different orientations of the dipole, i.e from small angles $\alpha$ (the dipole almost perpendicular to the dielectric) to $\alpha=\pi/2$ (dipole parallel to the dielectric sheet), and different dimensionless velocities $u= |\bf u| = v_x/(a\omega_{\rm pl})$. \\
\begin{figure}[!ht]
	\includegraphics[width=10cm]{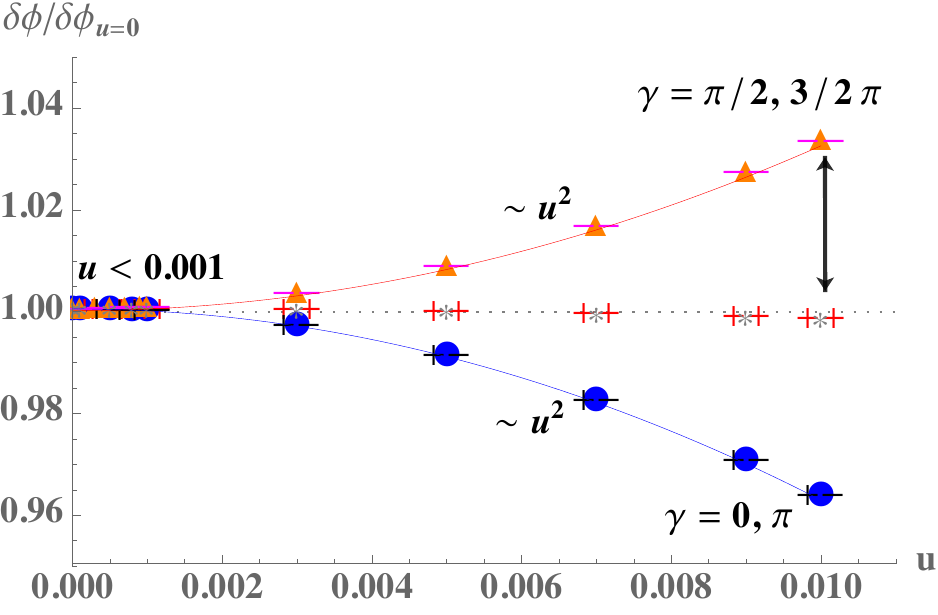}
	\caption{Ratio $\delta\phi /\delta\phi_{u=0}$  as a function 
		of the dimensionless tangential velocity $ {\bf u} = v_x/(a\omega_{pl}) {\bf\hat x}$ for different dipole orientations. 
		$\alpha = 0.1 $  is represented with the grey asterisks and is the less corrected case when $\gamma = \pi/2$. The rest of the cases shown have a dipole moment parallel to the 
		mirror ($\alpha = \pi/2$). Correction to the GP fits quadratically with $u$ as can be seen in Ref.\cite{EPL} for small tangential velocities. Parameters used: $\Delta/\omega_{\rm pl}= 0.9$, $\Gamma/\omega_{\rm pl} = 0.1$, 
		$\omega_0/\omega_{\rm pl}=0$, $N=15 $, $\theta_0=44.9^{\circ}$.}
	\label{Fig3}
\end{figure}

\noindent In Fig. \ref{Fig3} we show the correction to the phase $\delta\phi$ defined as the difference between $\phi_g$ the total geometric phase and the unitary phase $\phi_c$, i.e. 
$\delta \phi = \phi_g - \phi_c$, as a function of the tangential velocity of the particle for fixed initial angle $\theta_0$ and different orientations of the dipole. We can see that for very small velocities $u$ the GP is similar and comparable to the unitary one, since $\delta \phi = \delta \phi_{u=0} + \delta \phi_u$ has two contribution terms: one originated in the quantum field and dielectric sheet and another in the velocity of the atom. For very small values of $u$, the correction of the phase is mainly induced by the presence of the quantum field and dielectric sheet, which agrees with the threshold found for the appearance of the quantum frictional force at small velocities \cite{Farias2016}. However, when $u$ increases, the correction originated by the velocity becomes more important and the GP begins to show different behaviors as the dipole orientation is different. It is easy to see that when the dipole is perpendicular to the velocity and dielectric sheet, the correction to the GP does not vary significantly.  When the dipole is oriented parallel to the dielectric sheet, it can be found parallel to the direction of motion ($\gamma=0,\pi$) or perpendicular to this direction ($\gamma=\pi/2,3/2\pi$).  Once more, for very small values of the velocity there are no notorious differences. However, as the velocity $u$ grows in magnitude, we can infer that the dipole oriented in the direction of movement has the bigger correction. Since $u = v/(a\omega_{\rm pl})$ one can achieve relatively high values of $u$ just adjusting $a$ and/or $\omega_{\rm pl}$ but keeping $v$ under the non-relativistic range. Fig. \ref{Fig3} also shows that the correction to the GP fits quadratically with $u$ 
as can be seen in Ref.\cite{EPL} for small tangential velocities.

\begin{figure}[!ht]
	\includegraphics[width=10cm]{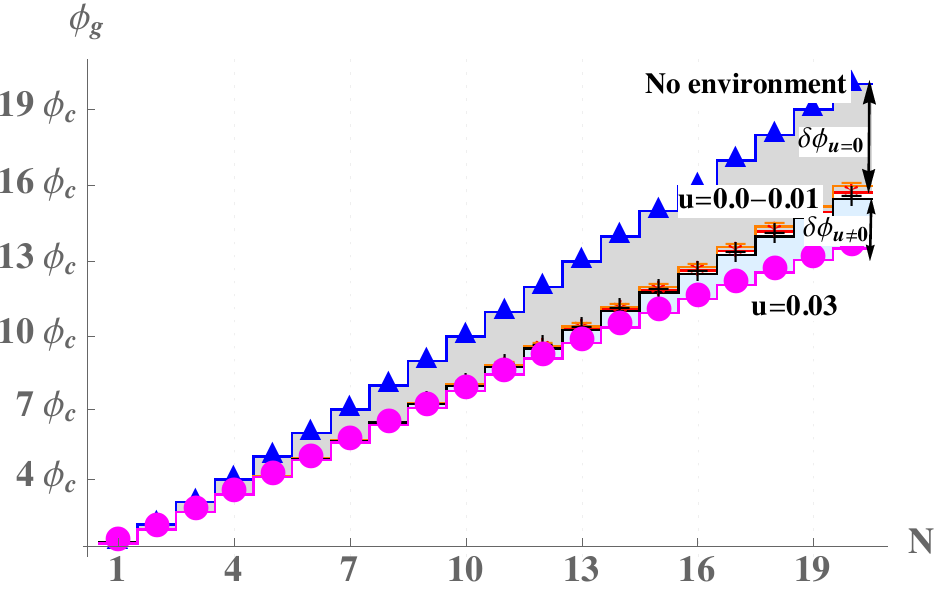}
	\caption{GP measured in units of the unitary GP ($\phi_c$) as function of time, in units of the 
		number of cycles of the isolated system ($N=t/\tau$). Red lines with triangle markers represent the phase acquired by the free evolution 
		of the particle. The remaining lines show the different corrections.  Parameters used: $\Delta/\omega_{\rm pl}= 0.9$, $\Gamma/\omega_{\rm pl} = 0.1$, 
		$\omega_0/\omega_{\rm pl}=0$, $\alpha=\pi/2$, $\gamma=0$, $\theta_0=44.9^{\circ}$.}
	\label{Fig4}
\end{figure}

\noindent 
The phase-lag induced by the motion on the fluctuating dipole results in dynamical effects (noise and dissipation in equation (\ref{master})). These corrections due to the quantum fluctuations can be evidenced as stochastic variations in the energy gap of the two-level system \cite{Buhmann}. In that case, this random variation produces noise effects on the internal degree of freedom that induce corrections in the GP of the two-level system.
We can further study the behavior of the GP as function of time  as shown in Fig.\ref{Fig4}. We plot the unitary GP that acquires the atom in each cycle with blue triangle markers  (a perfect straight line). However, this evolution is modified by the presence of the environment. 
For small velocities and few cycles, the GP acquired by the real system is similar to the unitary one. However, the lines start to differ considerably as time elapses and velocity increases. The correction is enhanced as the system evolves through more cycles, since the GP accumulates.
 The difference between the blue triangle markers and the lines with very small velocities (from $u=0$ to $u=0.01$) accounts mainly for the correction that suffers the GP when the two-level system evolves with vanishing velocity in the presence of a dielectric sheet $\delta \phi_{u=0}$, being negligible for few cycles but relevant later.  Likewise, the distance between the intermediate lines and pink circle markers is a strong evidence of the correction obtained when the velocity of the atom is of considerable importance $\delta \phi_{u\neq 0}$. 
 It is possible to see that for $N \gg 5$, the correction 
to the GP can be detected even for the smaller velocity $u$ considered. When $u= 0.03$, the correction 
for $N= 20$ is about $60\%$. \\

\noindent \textbf{ Experimental proposal}\\

\noindent Recalling the evolution described in Fig. \ref{Fig2} and the results obtained in Figs. \ref{Fig3} and \ref{Fig4}, we can assure this model is a good scenario for the measurement of the GP and its correction. In the case of a dipole orientation perpendicular to the dielectric sheet, the system stays ``robust" to the presence of the environment and the corrections to the geometric phase seem undetectable. However, in the opposite case, while the system preserves  purity for several cycles,  the correction induced by the  velocity becomes relevant at the same timescale, yielding the opportunity to detect traces of the velocity in the correction of the geometric phase.\\

\noindent 
A feasible experimental setup to perform an experiment to this end would be based on the use of a single NV center in diamond as an effective two-level system at the tip of a modified atomic force microscope (AFM) tip, which is able to maintain the distance between the NV center and the sample.  The distance can be controlled from a few nanometers to tenths of nanometers with subnamometer resolution.  The NV system  presents itself as an excellent tool for studying geometric phases. The NV center consists of a vacancy, or missing carbon atom, in the diamond lattice lying next to a nitrogen atom, which has substituted for one of the carbon atoms. 
The electron spin is the canonical quantum system and the NV center offers a system in which a single spin can be initialized, coherently controlled, and measured. It is also possible to mechanically move the NV center.

\begin{figure}[!ht]
\includegraphics[width=12cm]{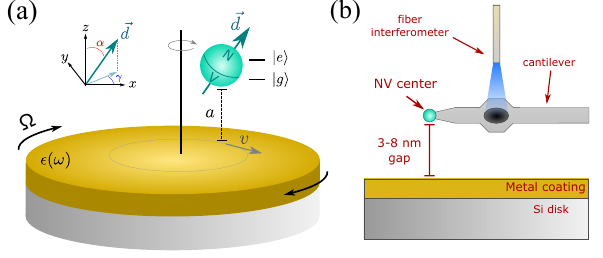}
\caption{(a) Schematic of the proposed experimental setup. A Au-coated Si disk rotates at angular velocity $\Omega$. The diamond NV center is placed at a distance $a$. The relevant coordinate system is as indicated in Fig.\ref{Fig1}(a). (b) Schematic of the setup. A diamond with an NV center is placed at the end of an AFM system. The AFM is used to keep the gap $a$ constant within 1 nm.}
\label{Fig5}
\end{figure}

\begin{figure}[!ht]
\includegraphics[width=10cm]{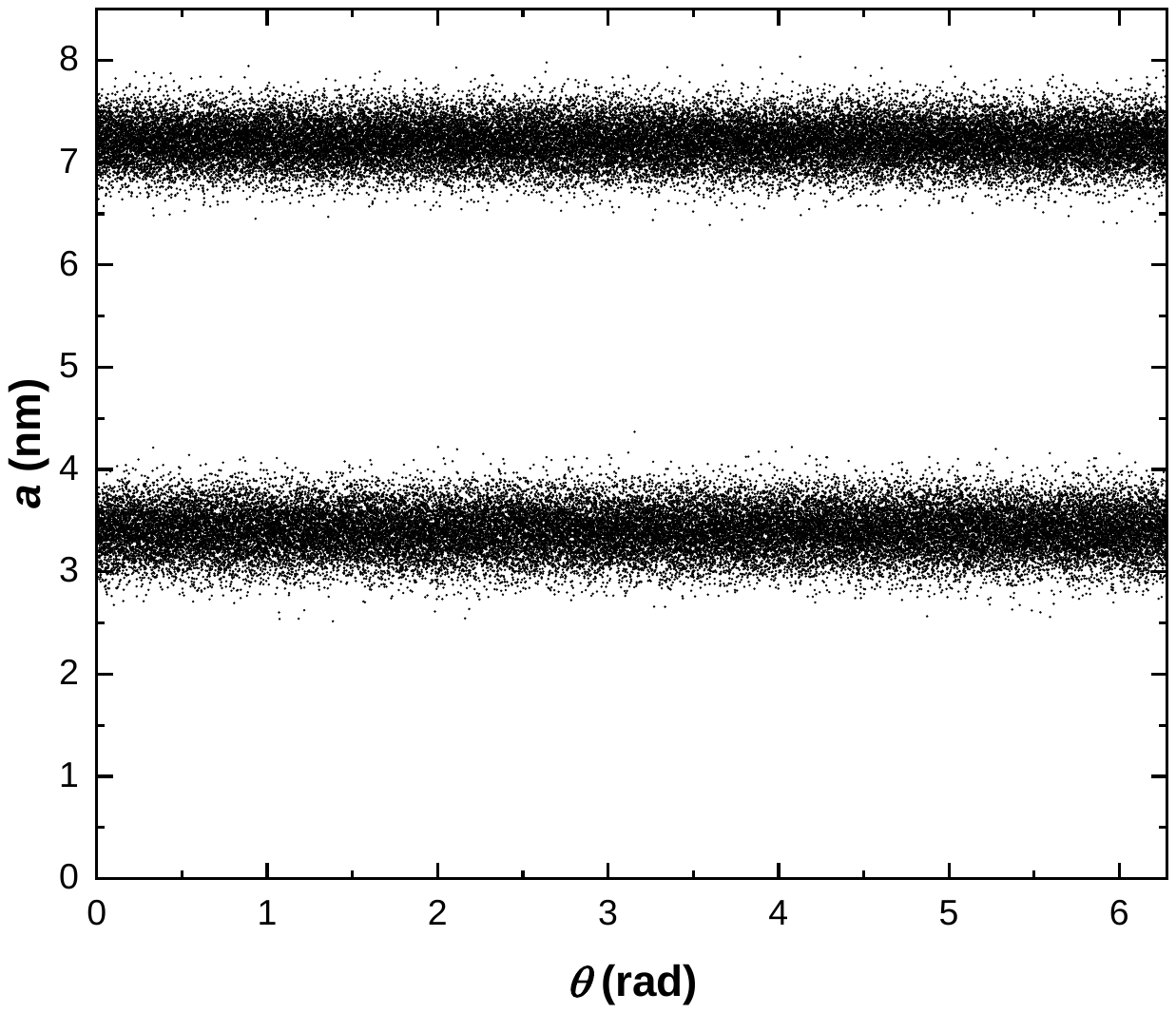}
\caption{Measurement of the separation between the AFM tip and a 12 cm diameter Au-coated Si disk rotating at $\Omega=2\pi 7000$Hz. The nominal separation $a$ between the tip and the sample are $7.2$ and $3.4$ nm. The AFM tip moves vertically $\approx 27.3$ nm to keep the separation constant.}
\label{Fig6}
\end{figure}

\noindent
In our proposed experimental setup, the sample is constituted by a Si disk laminated in metal (we propose to use Au or n-doped Si  coating. Parameters of the  Drude-Lorentz model for Au are $\omega_{\rm pl} = 1.37 \,10^{16}$rad/s; $\Gamma /\omega_{\rm pl} \sim 0.05$, and  $\omega_{\rm pl} = 3.5 \,10^{14}$rad/s; 
$\Gamma /\omega_{\rm pl} \sim 1$ for n-Si). The coated Si disk is mounted on a turntable, see Fig.\ref{Fig5} (a).  Although we are using a rotating table, non-inertial effects can be completely neglected in order to model a particle moving at a constant speed on the material sheet. Since it is critical to keep the separation uniform, to prevent spurious decoherence, it is important to asses the plausibility of the proposed experimental setup. We have checked that $12$cm  diameter Au-coated Si disks can be rotated up to $\Omega= 2\pi 7000$ rad/s. In these conditions, the measured wobble of the turntable is of the order of $10^{-8}$ radians (i.e. the vertical motion is 1nm at the edge of the disk).  While the overall change in thickness of the rotating plate could be as large as 50 nm at a given radius, the feedback control maintains the specified separation $a$ to better than $\delta a = 1$~nm. The experiment is doable at $a=3$ nm, with $\delta a$ (possible fluctuations in distance) induced decoherence effects being negligible compared to the quantum friction ones.  Fig. \ref{Fig5} (b) presents a schematic of the proposed experiment. 
We show in Fig. \ref{Fig6} the measured distance $a$ between the AFM tip and the rotating disk, indicating that the experiment is feasible. State-of-the-art phase-detection experiments in NV centers in diamond \cite{NVexp} permit the detection of $\sim$ 50 mrad phase change over $10^6$ repetitions. 

\begin{figure}[!ht]
\includegraphics[width=12cm]{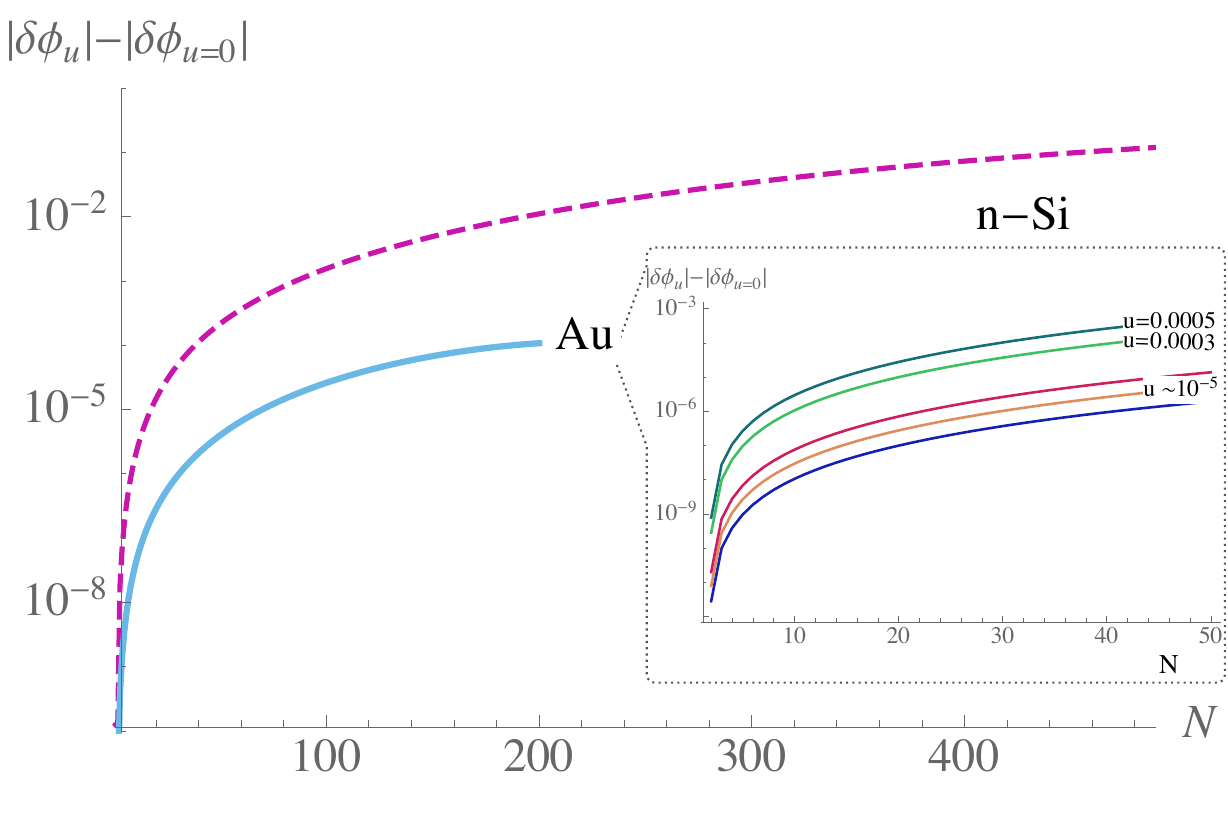}
\caption{The velocity-correction to the geometric phase is easily obtained by  $\vert \delta \phi_{u}\vert  - \vert  \delta \phi_{u=0}\vert $. As it has been reported, for very small velocities the presence of the quantum field dressed by the dielectric is dominant. However, for bigger velocities, the theory predicts that $\delta \phi_{u}$ becomes relevant making it possible to detect differences as the geometric phase accumulates. Figure shows numerical simulation in experimental conditions with n-Si  ($u = 0.0025$) and 
Au ($u = 6.4  \, 10^{-5}$) coating disk. The former case is measurable with the actual technology. The inset shows $\vert \delta \phi_{u}\vert  - \vert  \delta \phi_{u=0}\vert $ for a 
range of velocities $u$ derived by the assumption of values of $a$ ranged between 3 and 10 nm; all of them for the case of Au coating.}
\label{Fig7}
\end{figure}

\noindent 
Taking into account the experimental values considered in the proposal and the free parameters of the numerical simulations of the model, we can show that it is possible to detect a velocity dependence in the corrections of the geometric phase, as it accumulates in time. In Fig. \ref{Fig7}, we show the velocity dependent  correction to the geometric phase ($\delta \phi_{u\neq0}$). The greater the velocity achieved, the earlier in time the velocity-dependent corrections become relevant. In experimental conditions, we can achieve different velocities $u$ depending the metal coating of the  Si disk. When it is coated with n-doped Si, the dimensionless velocity $u$ is 
bigger, $u = 0.0025$ making it measurable with the actual technology. In the inset of \ref{Fig7} we plot $\vert \delta \phi_{u}\vert  - \vert  \delta \phi_{u=0}\vert $ 
for a range of velocities $\mathrm u$ derived by the assumption of values of $a$ ranged between 3 and 10 nm (as the one measured in Fig. \ref{Fig6}) for Au coating. \\

\noindent {\textbf{DISCUSSION}}\\

\noindent We have found a proper scenario  to indirectly detect the QF by measuring the GP acquired by a particle moving in front of a dielectric plate. We have shown that our system preserves purity for several cycles, which allows us to ensure that the GP could be measured. We have studied the correction to the unitary geometric phase and realized it can be decomposed in different contributions: on the one hand, a correction induced by the presence of the vacuum field dressed by the presence of a dielectric sheet and, on the other, a correction induced by the velocity of the motion of the particle in front of the dielectric sheet. We have also shown that after many cycles, the correction to the accumulated GP due to the velocity of the particle becomes relevant. It is important to remark that the mere presence of a velocity contribution in the noise corrections to the phase is an indication of the frictional effect over the quantum degree of freedom of the particle.  Finally, we have proposed an experimental setup which determines the feasibility of the experiment and would be the first one in tracking traces of quantum friction through the study of decoherence effects on a two-level system. The emerging micro- and nanomechanical systems promising new applications in sensors and information technology may suffer or benefit from non-contact quantum friction.  For this application, a better understanding  of non-contact friction is due and its detection one step forward.\\

\noindent {\textbf{METHODS}}\\

\noindent The coefficients appearing in the master equation (\ref{master}) are defined in terms of  
the noise and dissipation kernels, $\nu(t)$ and $\eta(t)$, respectively. 
The expressions for these kernels have been derived for the present case  and are written as
\begin{eqnarray}
\nu(t) =& -\frac{d^2\hbar}{\pi^2} \, \Gamma \omega^2_{\rm pl} \int_0^\infty \!\! d\omega \int_0^\infty \!\! dk\int_0^{2\pi} \!\! d\theta \frac{\omega k^2 e^{-2ka}}{
(\omega_0^2 - \omega^2)^2 + \omega^2\Gamma^2} \nonumber \\
&\times G(\theta, \alpha, \gamma) 
 \cos[(k \rm v \cos\theta - \omega) t], \nonumber \\
\eta(t)=& - \frac{d^2\hbar}{\pi^2} \,\Gamma \omega^2_{\rm pl} \int_0^\infty \!\! d\omega\int_0^\infty \!\! dk\int_0^{2\pi} \!\! d\theta \frac{\omega k^2 e^{-2ka}}{
(\omega_0^2 - \omega^2)^2 + \omega^2\Gamma^2} \nonumber \\
& \times G(\theta, \alpha,\gamma) \sin[(k \rm v \cos\theta - \omega) t],  
 \label{noisedisspkernels} 
 \end{eqnarray}
where we define $G(\theta, \alpha, \gamma) =\cos^2 \gamma \sin^2 \alpha \cos^2 \theta + \sin^2 \gamma \sin^ 2\alpha \sin^2 \theta 
+ 2 \cos\gamma \sin\gamma \sin^2\alpha \cos\theta \sin\theta + \cos^2\alpha$.
$\gamma$ and $\alpha$ are the spherical azimuthal and polar angles, respectively. In these coordinates, the 
components of the dipole moment are $d_x = d \cos\gamma \sin\alpha$, $d_y = d \sin\gamma\sin\alpha$, 
and $d_z = d \cos\alpha$ (see Fig.\ref{Fig1}(b)). The dipole moment can be written in terms of the static atomic polarization 
$\alpha_{\rm pol}$, $d^2 = 3/2 \hbar \Delta \alpha_{\rm pol}$.

\noindent The time dependent coefficients in equation (2) are finally written as

\begin{eqnarray} D(t) &=& \frac{1}{\hbar} \int_0^t ds\,\nu(s)\, \cos(2 \Delta s) \nonumber \\  
f(t) &=& -  \frac{1}{\hbar} \int_0^t ds\,\nu(s)\, \sin(2 \Delta s) \nonumber \\
\zeta(t) &=& -  \frac{1}{\hbar}  \int_0^t ds \, \eta(s)  \, \sin(2 \Delta s). \nonumber
\end{eqnarray}

\noindent The phase associated with the quasi-cyclic path traversed in state space  by the open system in a period $\tau$ have been  defined in \cite{pra1,prl} as 
\be \phi_g = \arg \left[ \sum_{k}
\sqrt{\epsilon_k(\tau) \epsilon_k(0)} \braket{k(0)}{k(\tau)}
e^{-\int_0^\tau dt \bra{k} \frac{\partial}{\partial t}\ket{k}
} \right] 
\label{gp} \nonumber
\ee
where $\ket{k(t)}$ and $\epsilon_k(\tau)$
are respectively the instantaneous eigenvectors and eigenvalues of $\rho (t)$. This is the main procedure done to obtain the geometric phase acquired with every set of parameters analyzed in the main text. 
\\

\noindent {\textbf{DATA AVAILABILITY}}\\

\noindent
The experimental data and the source code that support the findings of this study are available from the corresponding author upon reasonable request.\\

\noindent {{}

\noindent {\textbf{ACKNOWLEDGEMENTS}}\\

\noindent We acknowledge F. D. Mazzitelli for useful comments. We are supported by UBA, CONICET and ANPCyT--Argentina. R. S. Decca acknowledges support from the National Science Foundation through grants PHY-1607360 and PHY-1707985 and financial and technical support from the IUPUI Nanoscale Imaging Center, the IUPUI Integrated Nanosystems Development Institute, and the Indiana University Center for Space Symmetries. M. B. Farias acknowledges financial support from the national Research Fund Luxembourg under CORE Grant No. 11352881.\\

\noindent {\textbf{CONTRIBUTIONS}}\\

\noindent In balance all authors contributed equally to the work. F.L. proposed the idea. F.L. and M.B.F. performed the calculations of the model. The numerical simulations were carried out by A.S. and P.I.V. R.S.D. contributed to the experimental setup. All of us participated in the discussions and the writing of the manuscript.\\

\noindent {\textbf{COMPETING INTERESTS}}\\

\noindent The authors declare no conflict of interest.\\


\textbf{REFERENCES}}
\medskip
\begin{thebibliography}{9}

\bibitem{Casimir} Casimir, H.B.C. On the attraction between two perfectly conducting plates.
Proc. K. Ned. Akad. Wet. {\bf 51},793 (1948).
 
 \bibitem{Milonni} Milonni, P.W. The Quantum Vacuum (Academic Press, New York, 2003).

\bibitem{Bordag1} Bordag, M., Mohideen, U. and Mostepanenko, V.M. New Developments in the Casimir Effect. 
Phys. Rep. {\bf 353}, 1 (2001).

\bibitem{Bordag2}
Bordag, M., Klimchitskaya, G.L., Mohideen, U. 	and
Mostepanenko, V.M.  Advances in the Casimir Effect (Oxford University Press, Oxford, 2009).

\bibitem{Milton1} Milton, K.A. The Casimir Effect: Physical Manifestations of the Zero-
Point Energy (World Scientific, Singapore, 2001). 

\bibitem{Milton2}Milton, K.A. The Casimir effect: recent controversies and progress. J. Phys. A {\bf 37}, R209 (2004). 

\bibitem{Reynaud} Reynaud, S., Lambrecht, A., 
Genet, C., and Jaekel, M.T. Quantum vacuum fluctuations. C. R. Acad. Sci. Paris Ser. {\bf IV 2},
1287 (2001).

\bibitem{Lamoreaux} Lamoreaux, S.K. 
The Casimir force: background, experiments, and applications. Rep. Prog. Phys. {\bf 68}, 201 (2005).

\bibitem{DCE1} Dalvit, D.A.R., Maia Neto, P.A., and Mazzitelli, F.D. Fluctuations, dissipation and the dynamical Casimir effect. Lect. Notes Phys. {\bf 834}, 419 (2011).

\bibitem{DCE2}
Nation, P.D., Johansson, J.R., Blencowe, M.P.,  and Nori, F. Colloquium: Stimulating uncertainty: Amplifying the quantum vacuum with superconducting circuits. Rev. Mod. Phys. {\bf 84}, 1 (2012).

\bibitem{Pendry97} Pendry, J.B. Shearing the vacuum - quantum friction. J. Phys. Condens. Matter {\bf 9}, 10301 (1997).

\bibitem{debate1} Pendry, J.B. Quantum friction - fact or fiction?. New J. Phys. {\bf 12}, 033028 (2010).

\bibitem{debate2}  Pendry, J.B. Reply to comment on "Quantum friction - fact or fiction?".  New J. Phys. {\bf 12}, 068002
(2010).

\bibitem{debate3} Philbin, T.G.  and Leonhardt U.  No quantum friction between uniformly moving plates. 
New J. Phys. {\bf 11}, 033035 (2009).

\bibitem{debate4} Leonhardt, U.  Comment on "Quantum Friction - Fact or Fiction?". New J. Phys. {\bf 12}, 068001 (2010).

\bibitem{vp2007} Volokitin, A. I. and Persson, B. N. J. Near-field radiative heat transfer and non-contact friction. Rev. Mod. Phys. {\bf 79}, 1291 (2007).

\bibitem{Mohideen1} Lamoreaux, S. K. Erratum: Demonstration of the Casimir Force in the 0.6 to 6 $\mu$m Range. Phys. Rev. Lett. {\bf 78}, 5 (1997). 

\bibitem{Mohideen2} Mohideen, U. and Anushree Roy. Precision Measurement of the Casimir Force from 0.1 to 0.9 $\mu$m. 
Phys. Rev. Lett {\bf 81} 4549 (1998). 

\bibitem{Mohideen3} Ederth, T. Template-stripped gold surfaces with 0.4-nm rms roughness suitable for force measurements: Application to the Casimir force in the 20?100-nm range. Phys. Rev. A {\bf 62}, 062104 (2000). 

\bibitem{Mohideen4} Chan, H.V., Aksyuk, V.A., Kleiman, R.N., Bishop, D.J., and Capasso, F. Quantum Mechanical Actuation of Microelectromechanical Systems by the Casimir Force. Science {\bf 291}, 1941 (2001). 

\bibitem{Mohideen5} Bressi, G., Carugno, G., Onofrio, R., and Ruoso, G. Measurement of the Casimir Force between Parallel Metallic Surfaces. Phys. Rev. Lett. {\bf 88}, 041804 (2002). 
 
\bibitem{Mohideen6} Decca, R.S., L\'opez, D., Fischbach, E., and Krause, D.E. Measurement of the Casimir Force between Dissimilar Metals. Phys. Rev. Lett. {\bf 91}, 050402 (2003). 

\bibitem{Wilson} Wilson, C.M., et al. Observation of the dynamical Casimir effect in a superconducting circuit.
 Nature (London) {\bf 479}, 376 (2011).

\bibitem{Lahteenmaki} Lähteenmäki P., Paraoanu G.S., Hassel J., Hakonen P.J. Dynamical Casimir effect in a Josephson metamaterial. Proc. Natl. Acad. Sci U.S.A. {\bf 110}, 4234 (2013).

\bibitem{Gotsmann} Gotsmann B. Tribology: sliding on vacuum. Nature Materials, {\bf 10}, 87 (2011).

\bibitem{spheres} Zhao, R., Manjavacas, A., de Abajo, F.J.G. and Pendry, J.B. Rotational quantum friction. Phys. Rev. Lett. {\bf 109}, 123604(2012).

\bibitem{Farias1}Far\'\i as, M.B., Kort-Kamp, W.J. and Dalvit, D.A., 2018. Quantum friction in two-dimensional topological materials. Phys. Rev. B {\bf 97}, 161407(R) (2018). 

\bibitem{Farias2} Klatt, J., Far\'\i as, M.B., Dalvit, D.A.R. and Buhmann, S.Y. Quantum friction in arbitrarily directed motion. Phys. Rev. A {\bf 95}, 052510 (2017). 

\bibitem{Farias3} Marino J, Recati A, Carusotto I. Casimir forces and quantum friction from Ginzburg radiation in atomic Bose-Einstein condensates. Phys. Rev. Lett. {\bf118}, 045301 (2017).

\bibitem{intra} Intravaia, F., Oelschläger, M., Reiche, D., Dalvit, D.A.R. and Busch, K.. Quantum Rolling Friction. Phys. Rev. Lett., 123(12), p.120401 (2019). 

\bibitem{Fosco1} Far\'\i as, M.B., Fosco, C.D., Lombardo, F.C. and Mazzitelli, F.D. Quantum friction between graphene sheets. Phys. Rev. D {\bf 95}, 065012 (2017).

\bibitem{Fosco2} Far\'\i as, M.B., Fosco, C.D., Lombardo, F.C., Mazzitelli, F.D. and L\'opez, A.E.R. Functional approach to quantum friction: Effective action and dissipative force. Phys. Rev. D {\bf 91}, 105020 (2015). 

\bibitem{Buhmann} Klatt, J., Bennett, R. and Buhmann, S.Y. Spectroscopic signatures of quantum friction. Phys. Rev. A {\bf  94}, 063803 (2016).

\bibitem{Volokitin} Volokitin, A.I. and Persson, B.N.J.. Quantum friction. , Phys. Rev. Lett. {\bf 106}, 094502 (2011). 

\bibitem{vpPRB} Volokitin, A.I. and Persson, B.N.Y. Quantum Cherenkov radiation at the motion of a small neutral particle parallel to the surface of a transparent dielectric. Phys. Rev. B {\bf 94},  235450 (2016). 

\bibitem{EPL} Lombardo, F.C. and Villar, P.I. Geometric phase corrections on a moving particle in front of a dielectric mirror.  Europhys. Letts. {\bf 118}, 50003 (2017).

\bibitem{Berry}Berry, M.V. Quantal phase factors accompanying adiabatic changes. Proc. R. Soc. Lond. A {\bf 392} 45 (1984).

\bibitem{Fuentes1} Fuentes-Guridi, I., Carollo, A., Bose, S. and Vedral, V. Vacuum induced spin-1/2 Berry’s phase. Phys. Rev. Lett. {\bf 89} 220404 (2002).

\bibitem{Fuentes2} I. Fuentes Guridi, S. Bose, and V. Vedral, Phys. Rev. Lett. {\bf 85} 5018 (2000).

\bibitem{Martin} Fuentes-Guridi, I., Bose, S. and Vedral, V. Proposal for measurement of harmonic oscillator Berry phase in ion traps. Phys. Rev. Lett. {\bf 107} 131301 (2011). 

\bibitem{Martin2}  Mart\'in-Mart\'\i nez, E., Dragan, A., Mann, R.B. and Fuentes, I. Berry phase quantum thermometer. New Journal of Phys. {\bf 15}, 053036 (2013). 

\bibitem{Sjoqvist} Sj\"oqvist, E., Pati, A.K., Ekert, A., Anandan, J.S., Ericsson, M., Oi, D.K. and Vedral, V. Geometric phases for mixed states in interferometry. Phys. Rev. Lett {\bf 85}, 2845 (2000).

\bibitem{Moore} Moore, D.J. and Stedman, G.E. Non-adiabatic Berry phase for periodic Hamiltonians. Journal of Physics A {\bf 23}, 2049 (1990).

\bibitem{Tong1} Tong, D.M., Sj\"oqvist, E., Kwek, L.C. and Oh, C.H. Kinematic approach to the mixed state geometric phase in nonunitary evolution. Phys. Rev. Lett. {\bf 93}, 080405 (2004). 

\bibitem{Tong2} Tong, D.M., Sj\"oqvist, E., Kwek, L.C. and Oh, C.H., 2005. Erratum: Kinematic Approach to the Mixed State Geometric Phase in Nonunitary Evolution. Phys. Rev. Lett. {\bf 95}, 249902 (2005).

\bibitem{Zeilinger} Zeilinger, A., G\"ahler, R., Shull, C.G., Treimer, W. and Mampe, W. Single-and double-slit diffraction of neutrons. Rev. Mod. Phys. {\bf 60}, 1067, (1988).

\bibitem{Leek}Leek, Peter J., et al. Observation of Berry's phase in a solid-state qubit.
Science {\bf 318}, 1889 (2007).

\bibitem{rotating1} Maclaurin, D., Doherty, M.W., Hollenberg, L.C.L. and Martin, A.M., 2012. Measurable quantum geometric phase from a rotating single spin. Phys. Rev. Lett. {\bf 108}, 240403 (2012). 

\bibitem{rotating2} Wood, A.A., Hollenberg, L.C., Scholten, R.E. and Martin, A.M., 2019. Observation of a quantum phase from classical rotation of a single spin.  Phys. Re. Lett. {\bf 124}, 020401 (2020). 

\bibitem{dalvitNM} Intravaia, F., Behunin, R.O., Henkel, C., Busch, K. and Dalvit, D.A.R., 2016. Non-Markovianity in atom-surface dispersion forces. Phys. Rev. A {\bf 94}, 042114 (2016). 

\bibitem{barton} Barton, G. On van der Waals friction. II: Between atom and half-space. New J. Phys. {\bf 12}, 113045 (2010).

\bibitem{mastereq} Breuer H.P. and Petruccione F., {\it The Theory of Open Quantum Systems}, OUP Oxford, 2007. 

\bibitem{mastereq2} Benenti, G., Casati, G. and Strini, G. Principles of quantum computation and information: Volume II: Basic Tools and Special Topics. World Scientific Publishing Company. (2007)

\bibitem{refs1} Zurek, W.H., 2003. Environment-assisted invariance, entanglement, and probabilities in quantum physics. Rev. Mod. Phys. {\bf 75}, 715 (2003).

\bibitem{refs2} Lombardo, F.C. and Villar, P.I. Decoherence induced by a composite environment. Phys. Lett. A {\bf 336}, 16-24 (2005).

\bibitem{refs3} Villar, P.I. and Lombardo, F.C. Decoherence of a solid-state qubit by different noise correlation spectra. Phys. Letts. A {\bf 379}, 246-254 (2015).

\bibitem{pra1} Lombardo, F.C. and Villar, P.I. Geometric phases in open systems: A model to study how they are corrected by decoherence. Phys. Rev. A {\bf 74}, 042311 (2006).

 \bibitem{pra2} Lombardo, F.C. and Villar, P.I., 2008. Environmentally induced corrections to the geometric phase in a two-level system.  Int. J. of Quantum
Information {\bf 6},  707713 (2008).  

\bibitem{pra3} Villar, P.I., 2009. Spin bath interaction effects on the geometric phase. Phys. Lett. {\bf A} 373, 206 (2009). 

\bibitem{pra4} Villar, P.I. and Lombardo, F.C. Geometric phases in the presence of a composite environment. Phys. Rev. A {\bf 83}, 052121 (2011).

\bibitem{pra5} Lombardo, F.C. and Villar, P.I. Nonunitary geometric phases: a qubit coupled to an environment with random noise. Phys. Rev. A {\bf 87}, 032338 (2013). 

\bibitem{pra6} Lombardo, F.C. and Villar, P.I. Correction to the geometric phase by structured environments: The onset of non-Markovian effects. Phys. Rev. A {\bf 91}, 042111 (2015).

\bibitem{prl} Cucchietti, F.M., Zhang, J.F., Lombardo, F.C., Villar, P.I. and Laflamme, R. Geometric phase with nonunitary evolution in the presence of a quantum critical bath. Phys. Rev. Lett. {\bf 105}, 240406 (2010).


\bibitem{buhmannscheel} Scheel, S. and Buhmann, S.Y., 2009. Casimir-Polder forces on moving atoms. Phys. Rev. A {\bf 80}, 042902 (2009). 


\bibitem{nosqubit} Lombardo, F.C. and Villar, P.I. Corrections to the Berry phase in a solid-state qubit due to low-frequency noise. Phys. Rev. A {\bf 89}, 012110 (2014).

\bibitem{Farias2016} Far\'\i as, M.B. and Lombardo, F.C., 2016. Dissipation and decoherence effects on a moving particle in front of a dielectric plate. Phys. Rev. D. \textbf{93(6)}, 065035 (2016)


\bibitem{NVexp} Zhang, K., Nusran, N.M., Slezak, B.R. and Dutt, M.G., 2016. Experimental limits on the fidelity of adiabatic geometric phase gates in a single solid-state spin qubit. New J. Phys. {\bf 18} 053029 (2016).\\


\end{thebibliography}
\end{document}